\documentclass[twocolumn,groupedaddress]{revtex4}
\usepackage{graphicx}
\usepackage{amsmath}
\usepackage{braket}
\usepackage{setspace}
\usepackage{xcolor}
\usepackage{float}
\usepackage[cp1252]{inputenc}

\begin{document}


\title{Resistivity size effect due to surface steps on ruthenium thin films computed with a realistic tight-binding model}




\author{W. E. Richardson}
\author{E. R. Mucciolo}
\author{P. K. Schelling}
\altaffiliation{Advanced Materials Processing and Analysis Center, University of Central Florida}
\affiliation{Department of Physics, University of Central Florida, Orlando, Florida, 32826}
\email[]{patrick.schelling@ucf.edu}


\date{\today}

\begin{abstract}
A realistic tight-binding model is developed and employed to elucidate
the resistivity size effect due to steps on Ru thin films. The
resistivity of two different film orientations, $(0001)$ and $(1
\bar{1}00)$, is computed for transport along a $[1 1 \bar{2} 0]$
direction both for smooth surfaces and for surfaces with
monolayer-high steps. In the case of smooth films, the systems are
also studied using solutions to the Boltzmann transport equation
(BTE). Interestingly, the resistivity of $(1 \bar{1}00)$ surfaces
exhibits a significant size effect even in the absence of surface
steps. When monolayer-high steps are spaced $\sim 10$ nm apart, the
resistivity is shown to increase due to scattering from the
steps. However, only a small increase was found which cannot explain
the large effect seen in recent experiments with Ru thin films. This
highlights the need for further elucidation of the resistivity size
effect. Theoretical analysis suggest that films made from materials
with a relatively large ballistic conductance per area like Ru should
exhibit a reduced resistivity size effect. This result points to Ru as
a promising interconnect material. Finally, because a very efficient
algorithm for computing resistivity based on the kernel polynomial
method (KPM) is used, the approach fulfills a need for realistic
models that can span length scales directly relevant to experimental
results. The calculations described here include films approaching $5$
nm in thickness, with in-plane distances up to $\sim 160$ nm and
$3.8\times10^{5}$ atomic sites.

\end{abstract}


\maketitle


\section{Introduction}

The resistivity size effect, sometimes also called the classical size
effect, was first reported in 1901 \cite{thomson1901theory}. The
effect is observed as an increase in the resistivity $\rho$ of thin
metallic films with decreasing film thickness $d$. The
implications for interconnect technology have become significant as
feature sizes have decreased to nanometer scales
\cite{Barmak_2020}. It is generally accepted that diffusive surface
scattering of electrons is responsible for the classical size effect.
Theoretical descriptions often are based on the Fuchs-Sondheimer (FS)
model \cite{Fuchs_1938,Sondheimer:1952aa}. The FS model is obtained
from a solution of the Boltzmann transport equation (BTE) with
inelastic electron scattering occurring in the bulk and
partially-diffusive scattering at surfaces implemented through
boundary conditions. Another scattering mechanism occurs
via grain-boundary scattering in polycrystalline thin films. This is often
described using the Mayadas-Shatzkes model \cite{MayaShatz}. Yet efforts to understand and mitigate the
resistivity size effect have been hampered by an inadequacy of
theoretical modeling tools. Specifically, while it is accepted that
surface roughness due to various defects, including vacancies,
adatoms, and surface steps, are connected to diffusive electron
scattering, there has been a lack of theoretical approaches to
elucidate various mechanisms at the atomic scale and their relative
importance.

Recently, atomic-scale modeling of transport in thin films based on
first-principles electronic-structure methods has begun to address
mechanisms for diffusive surface scattering. The size-dependent
resistivity of thin Cu films with frozen-phonon disorder and missing
surface atoms has been explored using density-functional theory (DFT)
calculations employing the non-equilibrium Green's function (NEGF)
approach \cite{Zhou:2018aa}. There, it was determined, in general
agreement with previous theoretical expressions, that surface
scattering adds a term to the resistivity proportional to $1/d$, where
$d$ is the film thickness. The connection between the surface
roughness and the specularity parameter in the FS model was
theoretically explored for Cu films \cite{Ke:2009uf}. Using the same
NEGF approach, the influence of steps was determined for thin Cu films
with $(001)$ orientation \cite{Zhou_2018}. In those calculations, the
film thickness was $6$ monolayers (ML), corresponding to $d \approx
1.1$ nm. By computing electron transmission through a region with
steps of various heights, it was demonstrated that scattering from
steps results in an additional resistivity $\Delta \rho$ proportional
to $\omega/(\xi d)$, where $\omega$ is the root-mean squared film
roughness due to steps, and $\xi$ is the lateral correlation length of
the roughness. This general relation was experimentally verified for
tungsten thin films \cite{Zhou_2018}. However, while these studies
have shed significant light on the resistivity size effect, there
remains a need for realistic but more computationally efficient
approaches with the objective of elucidating transport over longer
length scales and thicker films more consistent with the experimental
range.

To address this need, we have developed an approach based on
tight-binding (TB) electronic structure calculations and the kernel
polynomial method (KPM) \cite{Weisse:2006aa} for enhanced numerical
efficiency. To approach the level of first-principles accuracy, the TB
model is parameterized by fitting to the results of DFT calculations
of ruthenium. Transport properties, namely, the resistivity, are
determined for Ru thin films with surface steps. The results are
analyzed using the approach implemented in Ref. \cite{Zhou_2018},
where the effect of scattering from steps was treated in the Landauer
formalism. The results demonstrate an additional resistivity $\Delta
\rho$ which is proportional to $1/d$, consistent with previous
theoretical and experimental results \cite{Zhou_2018}. However, the
model predictions indicate a significantly smaller effect than what is
predicted for Cu films and also seen in W films \cite{Zhou_2018}. This
is possibly due to a higher ballistic conductance per unit area for Ru
in contrast to Cu and W.

\section{Theory and Methodology}

\subsection{Model description and parameterization}

In the tight-binding (TB) approach, a local basis set is used with
hopping matrix elements described within the Slater-Koster formalism
\cite{Slater:1954aa}. The single-particle eigenstates $|\psi_{\lambda}
\rangle$ can then be specified as a linear combination of the
tight-binding basis states,
\begin{equation}
|\psi_{\lambda} \rangle = \sum_{i,\kappa} c_{\lambda, i \kappa} |\phi_{i \kappa} \rangle,
\end{equation}
where $\lambda$ represents the eigenstate, $i$ represents an atomic
site, and $\kappa$ represents one of the tight-binding orbitals on the
site. The expansion coefficients $c_{\lambda, i \kappa}$ can be
determined by direct diagonalization of the TB Hamiltonian. The
starting point for the TB model was taken from the model for Ru
developed in Ref. \cite{Mehl:1996aa,HbBSES-DAP}. However, in the model
developed here, we assume an orthogonal basis in contrast to the
non-orthogonal basis used in the original model. Consequently, the
model needed to be reparameterized using DFT results. As in the
original TB model \cite{Mehl:1996aa}, the DFT energies were shifted so
that the total energy was equal to the sum of the eigenvalues of the
occupied electronic states. This involves a shift in the eigenvalues
$\epsilon_{\lambda}$ by an energy $V_{0}$. The new eigenvalues are
then $\epsilon_{\lambda}^{\prime}= \epsilon_{\lambda}+V_{0}$ and the
total energy is
\begin{equation}
E_{\rm tot} = \sum_{\lambda} f(\epsilon_{\lambda}^{\prime} - \mu^{\prime} )
\epsilon_{\lambda}^{\prime},
\end{equation}
where $f(\epsilon)$ is the Fermi-Dirac distribution function at zero
temperature and $\mu^{\prime}$ is the electronic chemical
potential. The shift $V_{0}$ was made separately for each structure
simulated with DFT so that the computed cohesive energy of the fit TB
model would accurately reproduce the DFT results.  Finally, the TB
model includes 5s, 5p, and 4d orbital states, for a total of 9
orbitals per Ru site.

Parameterization of the model was accomplished by a detailed fit to
electronic band structures for bulk Ru obtained from the DFT
calculations. First-principles DFT results were obtained using Quantum
Espresso \cite{QE-2009,QE-2017,QE-2020}. The pseudopotential used
was obtained from Pslibrary 1.0.0 \cite{PPLib} and was generated via
the projector-augmented wave (PAW) method \cite{PAW94} and includes
contributions from 4s, 5s, 4p, 5p, and 4d orbitals. The
Perdew-Burke-Ernzerhof (PBE) exchange-correlation functional
\cite{PBE96} was used in the calculations. Self-consistent
calculations were performed using a primitive unit cell with a
20$\times$20$\times$20 Monkhorst-Pack mesh
\cite{Monkhorst:1976ta}. The cutoff energies used for the wave
function and density were 60 Ryd and 480 Ryd, respectively.

\begin{figure}
\includegraphics[width=\linewidth]{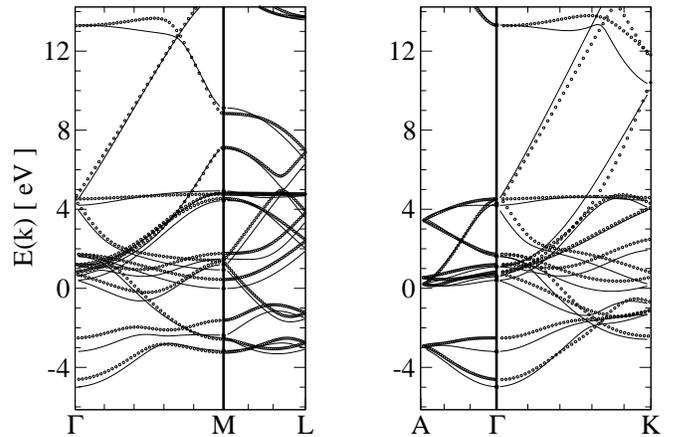}
\caption{Band diagram for ruthenium as calculated using DFT (solid
  lines) and the TB model (dots) for $a=2.74$~\AA\ and
$c=4.357$~\AA\ . The Fermi energy is located at
  $E_F=3.16$ eV.}
\label{fig:bands}
\end{figure}
While a description of the original TB model has been reported
elsewhere \cite{Mehl:1996aa}, some details on the dependence of the
on-site energies and hopping integrals are given in Appendix
A. Fitting to the DFT database was accomplished by a random walk in
parameter space using the original parameters as a starting point. At
each iteration, a small random change was made to each of the TB
parameters. This new trial set was used to build the TB
Hamiltonian and calculate the band structure for the atomic structure
and $k$-points specified in the training data. The training data
consisted of 7 hcp structures with different lattice parameters a and
c to capture the distance dependence in the hopping integrals and the
environmental dependence for the on-site energy terms. For each of the
7 structures, there were 40 total $k$-points sampling a path through
Brillouin zone both along high-symmetry directions and at
high-symmetry points. At each $k$-point, the character corresponding
to each eigenvalue was matched between the DFT and TB
calculations. The character table corresponds to the representations
of the space group P6$_3$mmc for the hcp lattice. We note that without
matching representations, the risk of misfitting the bands is
substantial, especially in cases in which the bands are very close
together in energy or in the case of band crossings. A cost was
assigned to the proposed parameterization by calculating the total sum
of squared differences between the DFT and TB energy bands at each
$k$-point and for each structure. In computing the cost, each band had
an associated weight which controlled how much influence that band had
on the final fit. The weights controlled the relative contribution
from each band to the total sum. Since this TB model was intended to
be used to calculate transport properties, the highest weights were
given to the bands in a region centered on the Fermi energy, extending
3 eV above and 3 eV below the Fermi energy, which will be referred to
as the conduction region. The low-lying, occupied bands in the region
below the conduction region were weighted about 20\% less than those
in the conduction region. Finally, the higher energy bands in the 10
eV to 30 eV range, which stay far away from the conduction region,
were weighted the least; up to 90\% less than those in the conduction
region.
In this way, the cost function favors fitting of the electronic states
responsible for electron transport. During each iteration of the fit,
the cost function was compared to the previous iteration. In cases
where the cost increased, the trial parameters were rejected.  If the
cost decreased, the trial parameters were accepted.  The iterative
minimization was allowed to proceed until a minimum value of the cost
function was obtained. The particular model obtained in this way
depends on the specific DFT data points used and the details of the
cost function. Moreover, the minimization procedure may not represent
a global minimum.

\begin{figure}
\includegraphics[width=\linewidth]{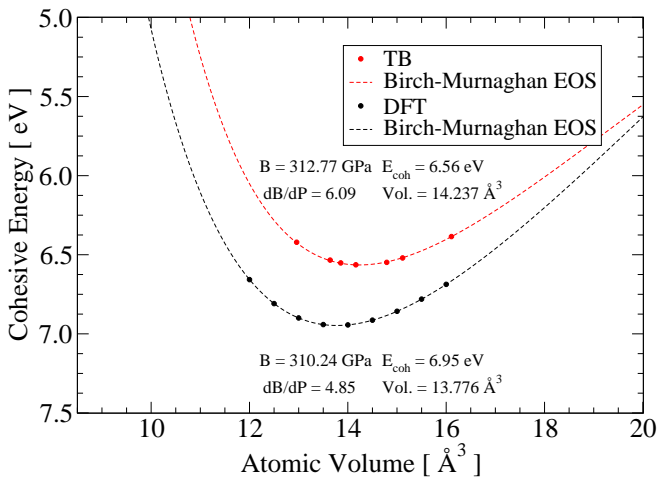}
\caption{Theoretical cohesive energy calculated by DFT (black) and TB
  (red). The Dashed lines correspond to the Birch-Murnhaghan equation
  of state fit to the respective data.}
  \label{fig:cohesive_energy}
\end{figure}

The resulting model showed quite good agreement for the electronic
band-structure, especially near the Fermi energy $E_{F}$. Sample
results corresponding to the TB ground-state structure are shown in
Fig. \ref{fig:bands} for $a=2.74$~\AA\ and $c=4.357$~\AA\ along a few
high-symmetry directions in the Brillouin zone. In addition to
excellent fits to the electronic band structure, the model is also
able to reproduce the cohesive energy over a range of atomic
volumes. In Fig. \ref{fig:cohesive_energy}, the cohesive energy
extracted from both the TB model and the DFT calculations are
shown. The specific points correspond to the cohesive energy for a
given volume with $c/a$ optimized. The resulting cohesive energy
curves were fitted with the Birch-Murnaghan equation of state
\cite{Birch:1947to,Murnaghan_1944} to obtain the atomic volume
$V_{0}$, cohesive energy $E_{\rm coh}$, bulk modulus $B$, and pressure
derivative $B^{\prime}$ of the bulk modulus. These values are shown in
Table \ref{tab:strucprop} along with a comparison to
experiment\cite{etde_7307886, KGJr-SSP, Wyck-CS} and the results
obtained from the original non-orthogonal TB model
\cite{Mehl:1996aa}. The results demonstrate good agreement between TB,
DFT, and experiment where results are available.

\begin{table}[ht]
\begin{tabular}{|l|c|c|c|}
 \hline
 $\qquad\qquad$ & $\quad$\textbf{Tight Binding}$\quad$ & \textbf{DFT}& \textbf{Experiment}\\
 \hline
 a (\AA) & 2.74  & 2.722   & 2.71 \\
             & (2.68)          &       &         \\
  \hline
 c (\AA) & 4.357 & 4.295  & 4.279  \\
 &  (4.26 )        &       &         \\
   \hline
 V$_0$ (\AA$^3$) & 14.24  &13.78 & 13.61 \\
 
  &   (13.25)     &      &         \\
     \hline
 E$_{\rm coh}$ (eV) &  6.56  & 6.95  & 6.62 \\
    \hline
 B (GPa) & 313 & 310  & 321 \\
&     (360)    &       &         \\
     \hline
 $\partial B/\partial P$ & 6.09 & 4.847 & N.A. \\
 \hline
\end{tabular} 
\caption{The structural properties of ruthenium determined with DFT
  and TB and compared to experimental values. Experimental data from
  \cite{etde_7307886, KGJr-SSP, Wyck-CS}. Wherever available, values
  reported from the original non-orthogonal TB model
  \cite{Mehl:1996aa} are given in parentheses.}
\label{tab:strucprop}
\end{table}

\begin{widetext}

\begin{figure}
  \includegraphics[width=0.75\linewidth]{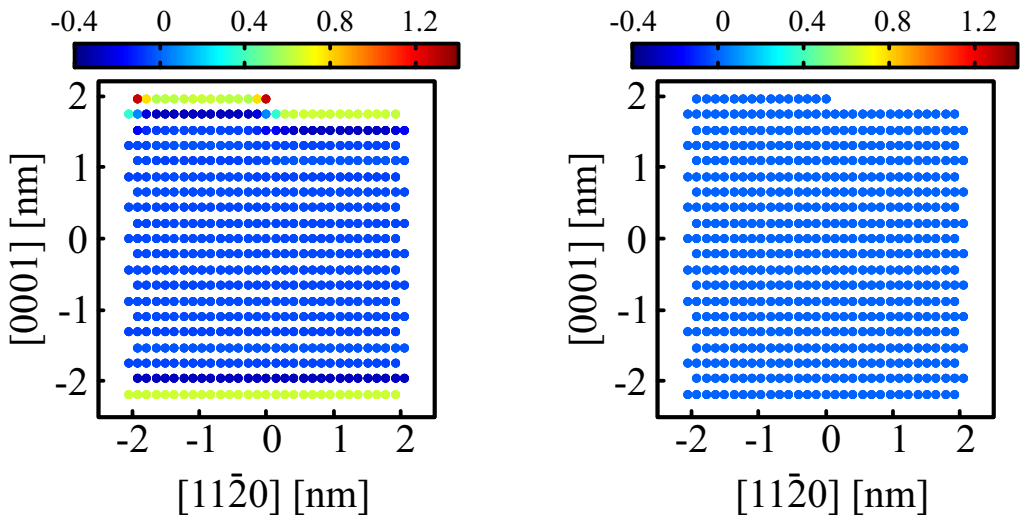}
  \caption{Occupation numbers $Q_{i}$ obtained before correction (left
    panel) and after correction (right panel) of the local site
    energies. On the right, it can be seen that the values of $Q_{i}$ are each close to zero.  The structure corresponds to a step on a $(0001)$
    surface.}
  \label{fig:occ}
\end{figure}

\end{widetext}

When simulating film structures, it was found using direct
diagonalization that under-coordinated surface sites have excess
electron site occupancy in comparison to bulk sites. We define the
excess occupancy of a site $Q_{i}$ using the TB expansion
coefficients,
\begin{equation}
Q_{i} = \sum_{\kappa , \lambda} f(\epsilon_{\lambda} -
\mu)\, c^{*}_{i\kappa,\lambda}\, c_{i\kappa,\lambda} - Z_{i},
\end{equation}
where $Z_{i}=8$ for each Ru site corresponding to the number of
valence electrons. For bulk Ru sites, $Q_{i}=0$. Results for $Q_{i}$
are shown in Fig. \ref{fig:occ} for a surface step in a Ru thin
film. While small deviations $Q_{i} \ne 0$ might be physical,
Fig. \ref{fig:occ} shows values which correspond to as much as one
additional electron per surface site, resulting in a rather large
negative surface charge density. Previous TB models of metallic
surfaces which have included Coulomb self-consistency tend to show
much smaller local site charges \cite{Erdin:2005vj}. For example, when
Coulomb interactions along with onsite energy terms are included,
deviations from $Q_{i}=0$ for the $(0001)$ surface of hcp Ti was found
to be much less that $\pm 0.01$. This reflects the substantial energy
cost associated with accumulation of charge on a few sites. To address
this issue without including self-consistent Coulomb interactions, we
implemented adjustments to surface site energies to enforce $Q_{i}=0$
at all sites. The results obtained with these corrections are shown in
Fig. \ref{fig:occ}. The approach used a local Lagrange multiplier for
each surface site which was chosen to exactly enforce $Q_{i}=0$. The
Lagrange multipliers were iteratively varied during the calculation
until $Q_{i}=0$ to within a desired threshold was obtained at each
site. Modifications to the local site energies were retained for
subsequent transport calculations.

In determining the site corrections, the structure displayed in Fig. \ref{fig:occ} was used with periodic boundary conditions in the
plane. Specifically, the structure in Fig. \ref{fig:occ} represents one repeat unit used to
determine the adjustments to onsite energies. The repeat unit was chosen to be long enough along the $[11\bar{2}0]$ so that adjacent steps did not interact. The site corrections obtained for surface sites and sites in the vicinity of a step were then applied to the much large film structures. An identical approach was applied to determine corrections for $(1\bar{1}00)$ films.

\subsection{Transport}

\indent Transport calculations were performed using KPM
\cite{Weisse:2006aa}. This was accomplished by expressing the
components of the conductivity tensor in terms of a double KPM
expansion in the energy domain. The components $\sigma_{\alpha
  \beta}(E)$ of the conductivity tensor are obtained from the
Kubo-Greenwood formula
\begin{equation}
\sigma_{\alpha\beta} \left(E\right) = \frac{2\pi\hbar e^2}{\Omega}
\text{Tr} \left[ \hat{v}_{\alpha}\, \delta(E-\hat{H})\,
  \hat{v}_{\beta}\, \delta(E-\hat{H}) \right],
\label{eqn:kubo1}
\end{equation}
where the system volume is $\Omega$ and $\hat{v}_{\alpha}$ is the
electron velocity operator defined using the Heisenberg
equation-of-motion,
\begin{equation}
\hat{v}_{\alpha} = \frac{d\hat{x}_{\alpha}}{dt} = \frac{i}{\hbar}
\left[\hat{H},\hat{x}_{\alpha}\right],
\label{eqn:heis}
\end{equation}
where the $\hat{x}_{\alpha}$ refer to the Cartesian components $\alpha$ of
the site coordinates. To efficiently evaluate the trace, we used the 
truncated-basis approximation which employs a set of $R$ 
$\left(9\times N\right)$ dimensional position-space random vectors $|R_{r}
\rangle$ to estimate the trace, namely,
\begin{equation}
\sigma_{\alpha\beta}\left(E\right) \approx \frac{2\pi\hbar
  e^2}{\Omega} {1 \over R} \sum_{r=0}^{R-1}\langle R_{r} |
\hat{v}_{\alpha} \delta(E-\hat{H}) \hat{v}_{\beta} \delta(E-\hat{H}) |
R_{r} \rangle.
\end{equation}
The length of the vector $|R_{r}\rangle$ therefore represents the $N$ sites with $9$ orbitals on
each site.
We define vectors
\begin{equation}
| \Psi_{\alpha, r}(E) \rangle = \delta(E-\hat{H}) \hat{v}_{\alpha} |
R_{r} \rangle
\end{equation}
and
\begin{equation}
|\Phi_{\beta,r}(E) \rangle = \hat{v}_{\beta}\delta(E-\hat{H}) | R_{r}
\rangle,
\end{equation}
which allow for a more compact expression for the conductivity tensor,
\begin{eqnarray}
\sigma_{\alpha\beta}\left(E\right) \approx \frac{2\pi\hbar
  e^2}{\Omega} {1 \over R} \sum_{r=0}^{R-1}\langle \Psi_{\alpha, r}(E)
|\Phi_{\beta,r}(E) \rangle.
\end{eqnarray}
To evaluate the vectors recursively for each random vector $|R_{r}
\rangle$, the operators are approximated using KPM with $N_{T}$ terms
in the expansion:
\begin{equation}
| \Psi_{\alpha, r}(E) \rangle = {1 \over \pi \sqrt{1-\varepsilon^{2}}}
\left[ g_{0}+ \sum_{n=1}^{N_{T}-1} g_{n}\, T_{n}(\varepsilon)\,
  T_{n}(\hat{h})\, \hat{v}_{\alpha} \right] |R_{r} \rangle
\label{eq:Psi_alpha}
\end{equation}
and
\begin{equation}
|\Phi_{\beta,r}(E) \rangle = {1 \over \pi \sqrt{1-\varepsilon^{2}}}
\left[ g_{0}+ \sum_{n=1}^{N_{T}-1} g_{n}\, T_{n}(\varepsilon)\,
  \hat{v}_{\beta}\, T_{n}(\hat{h}) \right] |R_{r} \rangle,
\label{eq:Psi_beta}
\end{equation}
where $T_n(z)$ are Chebyshev polynomials of the first kind and $g_n$
are kernel coefficients.
In Eqs. (\ref{eq:Psi_alpha}) and
(\ref{eq:Psi_beta}), the energy and the Hamiltonian have been rescaled
to fit within a unit interval in order to comply with the Chebyshev
polynomial's convergence range: $\varepsilon = (E-E_m)/\Delta E$ and
$\hat{h} = (\hat{H} - E_m)/\Delta E$, where $E_m$ and $\Delta E$ are
the band middle point and half width, respectively.

Because the Hamiltonian ${\hat H}$ in a tight-binding basis is sparse,
evaluating the conductivity tensor in this way scales linearly with
system size for a fixed number of random vectors $R$. While the
number of random vectors $R$ or terms $N_{T}$ required for 
convergence of the conductivity tensor may increase with the system
size, often this dependence is sublinear, resulting in an overall
scaling which is quite advantageous with respect to exact
diagonalization.

All KPM conductivity calculations were done using the Jackson kernel
and $N_{T}=242$ terms in the Chebyshev expansions. 
Explicitly, the coefficents $g_n$ defined by the Jackson 
kernel take the functional form:
\begin{eqnarray}
g_n & = & \frac{1}{R+1} \left[ \left(R-n+1\right) \cos{\left(\frac{\pi
      n}{R+1}\right)} \right. \nonumber \\ & & \left.
  +\ \sin{\left(\frac{\pi n}{R+1}\right)}
  \cot{\left(\frac{\pi}{R+1}\right)} \right].
\end{eqnarray}
For ballistic or quasi-ballistic systems, in the absence of a relaxation time scale
such as elastic or inelastic scattering times, the number of terms in
the Chebyshev expansion affects the computed conductivity, with the
conductivity increasing approximately linearly with the number of
moments $N_{T}$ used in Eqs. \eqref{eq:Psi_alpha} and
\eqref{eq:Psi_beta}. Because the objective was to compare to
experimental results, $N_{T}$ was chosen everywhere to reproduce the
bulk room-temperature resistivity for transport in the basal
plane. Thus the truncation of the expansion acts similarly to an
inelastic relaxation process, mixing states within an energy window of
$\Delta E/N_T$ \cite{Ferreira2015}.

Bulk KPM calculations were done on a rhombohedral supercell containing
327680 atoms which measured approximately 17.5 nm on each side. Bulk
BTE calculations were done using a 2-atom primitive cell over a
64$\times$64$\times$40 $k$-point mesh.  This particular $k$-point mesh
was chosen to replicate an equivalent sampling of the Brillouin zone
for both the BTE and KPM calculations. The Cartesian $z$-axis was
chosen to lie along the crystallographic $[0001]$ direction, and the
Cartesian $x$-axis was chosen to lie along the $[1 1\bar{2} 0]$
direction. This places the basal plane of the crystal in the
xy-plane. As a result, there are only two unique, non-zero components
in the conductivity tensor.  For the $\sigma_{xx}$ and $\sigma_{yy}$
components of the conductivity tensor, 36 random vectors were
sufficient for convergence and for the $\sigma_{zz}$ component, 108
random vectors were used. Calculations of the energy-dependent
conductivity were tested for convergence relative to the random basis
by conducting a t'-test at each energy. Results were considered
converged when increasing the number of random vectors used to
calculate the stochastic trace yielded statistically indistinguishable
results for each energy. The conversion from conductivity to
resistivity was straightforwardly done by inverting the (diagonal)
conductivity tensor.

For KPM calculations of thin films, orthorhombic supercells were
generated with fixed length and width and varying
thicknesses. Transport was computed always in the plane of the film
along the $[11\bar{2}0]$ direction.  The length along $[11\bar{2}0]$
was approximately 160 nm.  The width of the films perpendicular to the
transport direction was approximately 5.7 nm for $\left(0001\right)$
films and 5.2 nm for $\left(1\bar{1}00\right)$ films. For BTE
calculations on smooth films, orthorhombic unit cells were generated
for each film thickness. To ensure the sampling of the Brillouin zone
was commeasurate with the KPM calculations, transport calculations
using BTE were done over a 584$\times$12$\times$0 $k$-point mesh.  For
all film calculations, the Cartesian $z$-axis was chosen perpendicular
to the surface, and the Cartesian $x$-axis was aligned with the
$[11\bar{2}0]$ direction.  Steps on thin films were generated by
removing or adding one layer of atoms from the top surface of a given
film, with the step perpendicular to the transport direction. For a
section of film with a given thickness, step lengths were pulled
randomly from a normal distribution centered on the intended
correlation length. For this study, steps with a correlation length of
10 nm were generated with a 5 nm standard deviation. For each film
thickness, an ensemble of stepped surfaces were generated for transport
calculations. The convergence of calculations for films with steps were
determined by comparing the standard error for the ensemble
calculation to the standard error for the random-vector truncated
basis. When the standard error of the ensemble was of the same order
of magnitude or smaller than the standard error introduced by the
truncated basis, the results were considered converged. For the
surface disorder examined in this study, it was found that a
relatively small ensemble consisting of five disorder realizations for
each thickness was sufficient to meet the conditions for
convergence. For all thin-film KPM calculations, 36 random vectors
were used to calculate the stochastic trace. Finally, the error bars
shown for smooth film KPM calculations reflect the uncertainty
introduced by the truncated random vector basis used for the
stochastic trace. For the stepped-film KPM calculations, the error bars
shown reflect both the combined uncertainty of the ensemble averaging
and that of the truncated random vector basis.

\begin{figure}
\includegraphics[width=\linewidth]{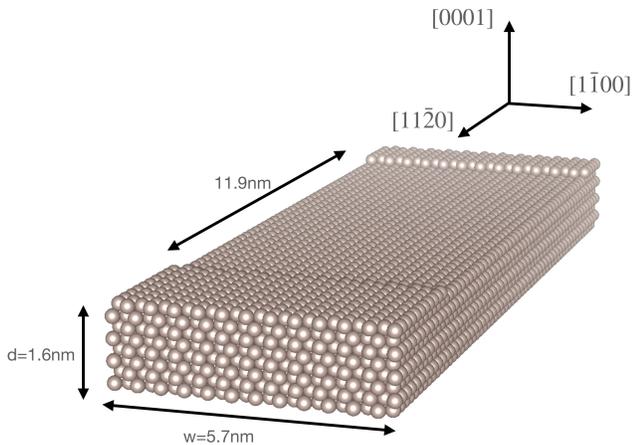}
\caption{Representative segment of a stepped $(0001)$ film. The film thickness $d=1.6$nm and width $w=5.7$nm are indicated.
The separation between step edges of $11.9$nm is consistent with $\xi=10$nm. Crystallographic directions also shown,
with periodic-boundary conditions applied in the $[1 \bar 1 00]$ and $[1 1 \bar{2}0]$ directions. }

\label{fig:film1}
\end{figure}

\begin{figure}
\includegraphics[width=\linewidth]{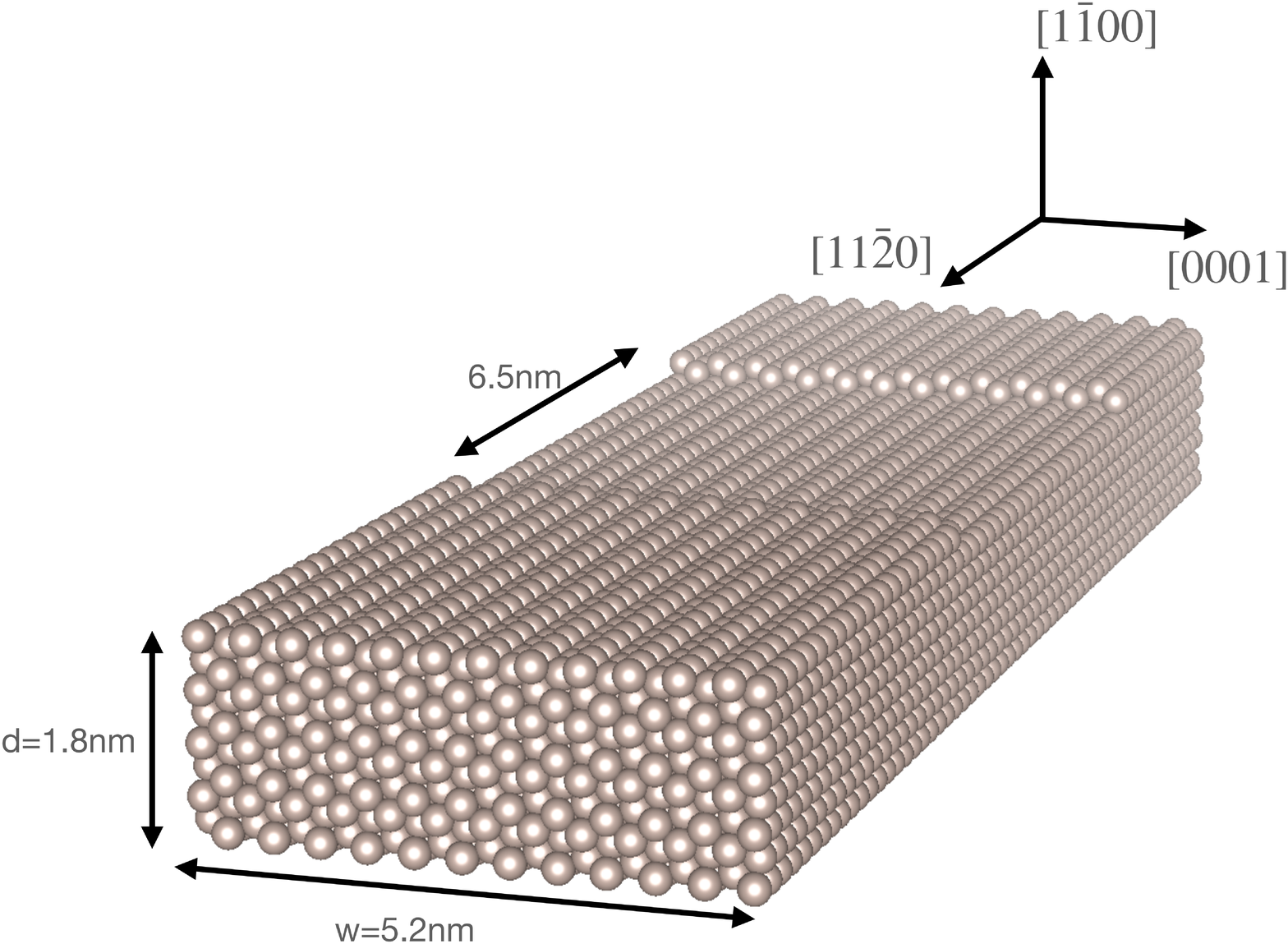}
\caption{Representative segment of a stepped $(1 \bar{1} 00)$ film. The film thickness $d=1.8$nm and width $w=5.2$nm are indicated.
The separation between step edges of $6.5$nm is consistent with $\xi=10$nm. Crystallographic directions also shown,
with periodic-boundary conditions applied in the $[0001]$ and $[1 1 \bar{2}0]$ directions. }

\label{fig:film2}
\end{figure}

Short representative segments of stepped films are shown in Fig. \ref{fig:film1} and Fig. \ref{fig:film2} for the $(0001)$ and $(1\bar 1 00)$ surfaces 
respectively. The figures include labels for the thickness $d$ and width $w$. The distance between
the steps is also indicated. The distance between the steps is comparable to the average step correlation length
$\xi=10$nm in both cases. Finally, periodic boundary conditions were applied in the plane, with the width $w$ the periodic
repeat distance shown for the direction perpendicular to the $[11\bar{2} 0]$ transport direction.

\section{Results}
 
\begin{figure}
\includegraphics[width=\linewidth]{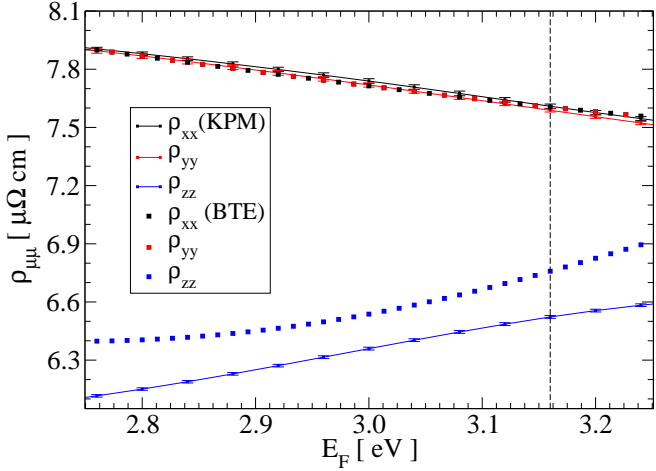}
\caption{Bulk resistivity calculated using Kubo-Greenwood via KPM
  (lines with error bars) and Boltzmann transport equation (points)
  via direct diagonalization. The bulk Fermi energy was found to be
  3.16 eV (indicated by a dashed line) and the film Fermi energies
  were all were between 2.9 eV and 3.2 eV. The black, red, and blue
  plots are $\rho_{xx}$, $\rho_{yy}$, and $\rho_{zz}$,
  respectively. In these plots, the x-axis lies along the $\left[1 1
    \bar{2} 0\right]$ direction and the z-axis lies along the
  $\left[0001\right]$ direction. }

\label{fig:bulk}
\end{figure}

The results for bulk resistivity determined using KPM are presented in
Fig. \ref{fig:bulk}. Past experiments
\cite{Powell67,Milosevic_2018,Ezzat_2019} report the room temperature,
basal plane resistivity for single crystal ruthenium as
$7.6\,\mu\Omega\,$cm and the c-axis resistivity to be
$5.8\,\mu\Omega\,$cm. As mentioned previously, the number of
moments $N_{T}$ in the KPM calculation was chosen to exactly reproduce
the basal plane resistivity. The c-axis resistivity of
$6.52\,\mu\Omega\,$cm obtained from KPM calculations is larger than
experiment yet still reasonable. For comparison, bulk transport was
also determined using the Boltzmann-Transport Equation (BTE) within
the single-relaxation time approximation. The elements of the BTE
conductivity tensor are computed using the expression
\begin{eqnarray}
\sigma_{\mu\nu} & = & \frac{2e^2\tau}{N_k V_{cell}}\sum_{\lambda}
\sum_{\vec{k}} \left(\vec{e}_{\mu}\cdot\vec{v}_{\lambda\vec{k}}\right)
\left(\vec{e}_{\nu}\cdot\vec{v}_{\lambda\vec{k}}\right) \nonumber \\ & & \times
\left(-\frac{\partial f}{\partial \epsilon}\right)_{\epsilon =
  E_{\lambda}\left(\vec{k}\right)},
  \label{BTE}
\end{eqnarray}
where $\partial f / \partial \epsilon$ is the derivative of the Fermi
function. The relaxation time $\tau=9.45$ fs was chosen in the BTE
calculations to also reproduce the experimental room-temperature
basal-plane resistivity $\rho_{RT}=7.6\, \mu \Omega\,$cm. The BTE
calculations were done with an electronic temperature $T_{e} = 2200$~K
to provide sufficient smearing for the electron occupations.
Fig. \ref{fig:bulk} also shows the results of BTE calculations within
the single relaxation-time approximation, which exhibits excellent
agreement with the KPM calculation although resistivity along the
$c$-axis is slightly overestimated.

\begin{figure}

\includegraphics[width=\linewidth]{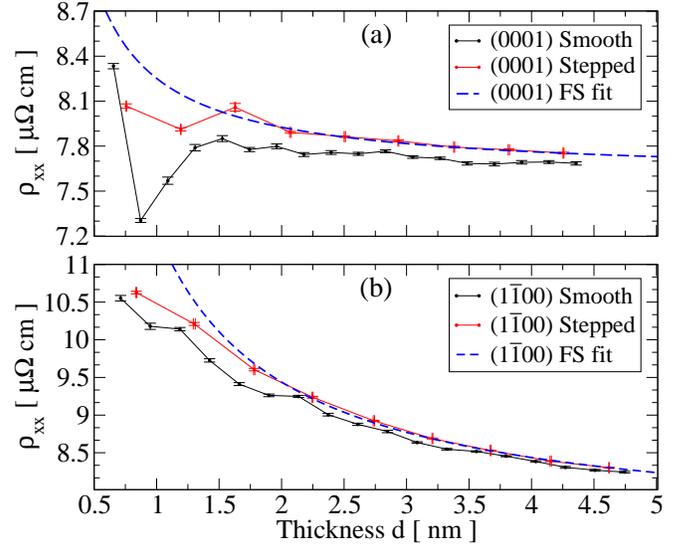}
\caption{The KPM resistivity calculations for the (a)
  $\left(0001\right)$ surface and (b) $\left(1\bar{1}00\right)$
  surface. Results for both smooth and stepped surfaces are plotted, with
  the simplified FS model fit to each stepped surface. }
  \label{fig:kpmfilms}
\end{figure}

Thin film KPM resistivity calculations are shown in
Fig. \ref{fig:kpmfilms}. Somewhat surprisingly, even films without
steps show increased resistivity with decreasing thickness, despite a
lack of diffusive surface scattering. This behavior is especially
pronounced for $\left(1\bar{1}00\right)$ films. Resistivity averaged
over an ensemble of stepped surfaces is also shown in
Fig. \ref{fig:kpmfilms}, demonstrating an increased resistivity due to
scattering from steps.
 
The FS model is given by the expression,
\begin{equation}
\rho = \rho_{0} \left[ 1-{3 \over 2 \kappa} (1 - p) \int_{1}^{\infty}
  \left({1 \over t^{3}}-{1 \over t^{5}} \right) {1-e^{-\kappa t} \over
    1-pe^{-\kappa t}} dt \right]^{-1},
 \label{fs1}
 \end{equation}
in which $\rho_{0}$ is the bulk resistivity, $d$ is the film
thickness, $\lambda$ is the inelastic mean-free path,
$\kappa=d/\lambda$, and $p$ is the specularity parameter.  The data
for stepped surfaces in Fig. \ref{fig:kpmfilms} was fit to the
simplified Fuchs-Sondheimer model. Specifically, the data in Fig. \ref{fig:kpmfilms}
for stepped films is fit using the expression,
\begin{equation}
\rho = \rho_{0} \left[ 1+ \left({3 \over 8}{\lambda (1-p) \over d}
  \right) \right].
\label{fs2}
\end{equation}
When surface scattering is entirely specular, $p=1$. By contrast, the
diffusive scattering limit corresponds to $p=0$.
The simplified FS model fit shown in Fig. \ref{fig:kpmfilms} was the
result of a two parameter fit, with $\rho_{0}$ as one fitting
parameter and the value of the product $\lambda(1-p)$ as the second
fitting parameter. For the $(0001)$ films, $\rho_{0}=7.62$
$\mu\Omega\,$cm and $\lambda(1-p)=0.208$nm were obtained from the
fit.  The much larger resistivity size effect computed for the $(1
\bar{1}00)$ films resulted in fit parameters $\lambda(1-p)=1.44$nm and
$\rho_{0}=7.44$ $\mu\Omega\,$cm. Interestingly, however, much of
the thickness dependence is seen even in the absence of steps and
hence without any diffusive scattering. This is most notable for the
$(1 \bar{1}00)$ films.
Increased resistivity is observed for films with both smooth and stepped
surfaces, with the stepped surfaces seeing a small additional
enhancement in resistivity. Of the studied surface configurations, the
resistivity-thickness relationship is most pronounced in
$\left(1\bar{1}00\right)$ oriented surfaces.  The coordination of
atoms at the film surfaces is the primary difference between the
$\left(0001\right)$ and $\left(1\bar{1}00\right)$ oriented smooth
films. The $\left(1\bar{1}00\right)$ surface is populated by sites
with two different coordination configurations, with some having $8$
nearest neighbors and the others having $10$ nearest neighbors. In
contrast, the $\left(0001\right)$ oriented surface is populated by
sites with only one coordination configuration, with each site having
$9$ nearest neighbors. The higher-coordinated $\left(0001\right)$
surface shows a weaker 1/d trend.

\begin{figure}
\includegraphics[width=\linewidth]{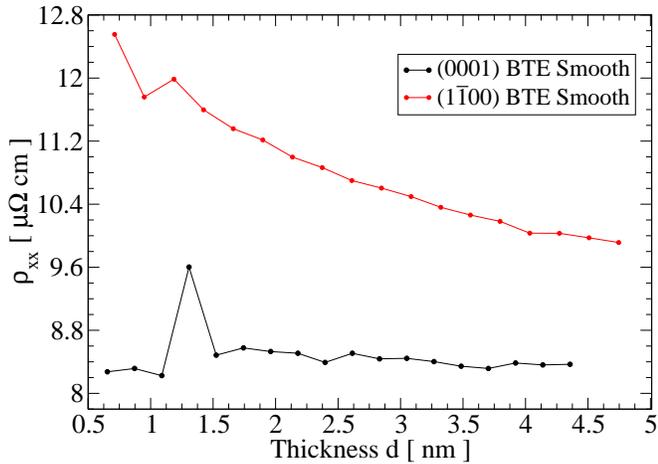}
\caption{Thin film resistivity for two smooth surfaces calculated using
  the Boltzmann transport equation. Both data sets follow the same
  trend as the smooth surfaces with resistivity calculated using KPM.}
\label{fig:btefilms}
\end{figure}

To verify that the resistivity size effect found for perfect films
without steps is due simply to differences in the Fermi surface, smooth
films were also studied using the BTE.
Since the BTE is only dependent upon the nature of the conducting
energy bands, any observed increases in resistivity as film thickness
decreases can be attributed to band structure effects. BTE
calculations of the resistivity are presented in
Fig. \ref{fig:btefilms}. The calculated resistivity for the smooth
$\left(1\bar{1}00\right)$ oriented surface reveals a fairly strong
dependence upon film thickness whereas the results for the
$\left(0001\right)$ oriented film shows a very weak dependence upon
film thickness. This follows the same general trend seen in
Fig. \ref{fig:kpmfilms} for KPM calculated thin film resistivity. This
establishes the validity of the KPM calculations and demonstrates that
the resistivity size effect can occur in the absence of diffusive
electron scattering simply due to changes in the electronic structure
due to surfaces. It can be seen, however, that the BTE calculations
produce somewhat higher resistivity values in the thin films and
appear to extrapolate to a higher resistivity in the limit of a bulk
film $d \rightarrow \infty$. This is most likely due to the finite
width of the computed films, and differences in how ``smearing''
occurs across the Fermi surface in the KPM and BTE calculations.  In
the BTE calculations, smearing is accomplished by a Fermi function
(see Eq. \ref{BTE}) and a finite electronic temperature $T_{e} =
2200$~K. However, despite the offset in resistivity, the dependence on
thickness $d$ is quite similar between KPM and BTE calculations.

\begin{figure}
\includegraphics[width=\linewidth]{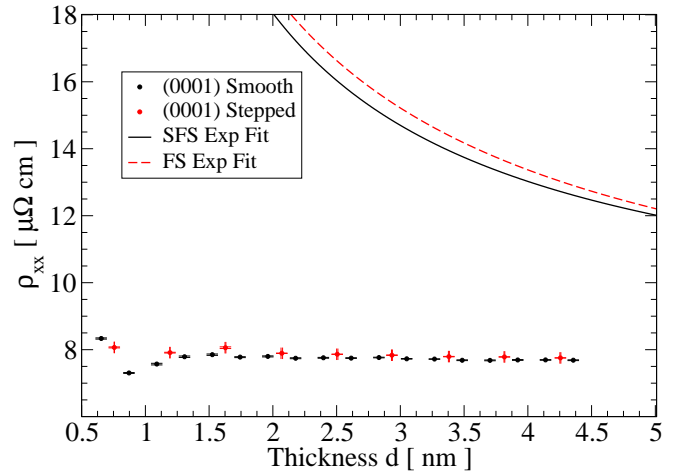}
\caption{The Fuchs-Sondheimer (FS) and Simplified Fuchs-Sondheimer
  (SFS) fits to the experimental data in \cite{Milosevic_2018} are
  shown above, respectively Eqs. \eqref{fs1} and \eqref{fs2}. The
  monoatomic steps alone cannot account for the resistivity size
  effect seen in \cite{Milosevic_2018}.}
\label{fig:expvskpm}
\end{figure}

The increase in resistivity from monolayer steps cannot account for
the experimentally-observed resistivity size effect. This is
illustrated in Fig. \ref{fig:expvskpm}, where the KPM calculated
resistivity for smooth films is compared to the FS fits to experimental
data obtained for single-crystal Ru $(0001)$ films reported in
\cite{Milosevic_2018}. In the $(0001)$ Ru films used in experiment,
atomic-force micographs were consistent with monolayer-high
steps. However, surface mounds were observed with widths in the range
$100-300$nm which might indicate that the step density of the
structures used in KPM calculations was actually greater than that of
the films studied in experiment \cite{Milosevic_2018}.  The parameters
used in the FS fits to experiment were $\rho_0=7.6 \mu\Omega\,$cm,
$\lambda=6.71$nm, and $p=0$ for corresponding to completely diffuse
scattering \cite{Milosevic_2018}.  Comparison of the KPM results to
experimental results for $(0001)$ films suggests that monolayer-high
steps do not result in enough diffuse scattering to account for the
classical size effect observed in experiment. In fact, it was noted by
the authors of the experimental study, using theoretical insight
obtained in a previous study\cite{Zhou_2018}, that steps would likely
contribute less than 1$\%$ of the observed resistivity size effect
\cite{Milosevic_2018}. The results computed using KPM agree with this
assessment. This indicates the likelihood of other diffuse scattering
mechanisms, or potentially a combination of different mechanisms.  We
will return to this point in the Conclusions section.

Despite the small effect, the increases to the resistivity due to
monoatomic steps can be quantified and interpreted based on the
analysis approach recently applied to describe experimental thin film
resistivity\cite{Zhou_2018}. This approach is based on the Landauer
formalism applied to electron scattering from steps. The details of
this approach are developed in Sec. \ref{sec:analysis}. By comparing
the resistivity of smooth films of a given thickness to a stepped film of
the same thickness, the rather small classical size effect seen in the
KPM calculations can be understood.

\section{Analysis}
\label{sec:analysis}

We analyze the transport results following the approach described
previously in Ref. \cite{Zhou_2018}. The basic idea is to treat
scattering from step edges within the Landauer formalism, such that
each step has a transmission probability $\eta(s)$ which depends on
the step height $s$. The resistance $\rho$ of a film with cross
sectional area $A=wd$, where $w$ is the width, is comprised of a
``flat film'' resistivity $\rho_{\rm ff}$ and an additional
resistivity term due to scattering from $N$ steps which occur along a
length $L$ of material,
\begin{equation}
\rho = \rho_{\rm ff} + {1 \over g_{0}L} \sum_{i=1}^{N} \left( {1 \over
  \eta_{i}} -1 \right).
\end{equation}
The term $\rho_{\rm ff}$ is due to scattering from point defects and
phonons, and represents the resistivity of a ``smooth'' film (i.e.,
without steps). The specific ballistic conductance $g_{0}$ is defined
as the ballistic conductance per cross-sectional area $A=wd$,
\begin{displaymath}
g_{0}={2 e^{2} \over h} {M \over w\, d},
\end{displaymath}
where $M$ is the number of conductance channels which scales linearly
with the cross sectional area.

Using NEGF methods \cite{Zhou_2018}, it
was found that the transmission coefficient for a step of height $s$
(including both an ``up'' and ``down'' step) in a film of thickness
$d$ is given by
\begin{equation}
\eta(s) = 1 - {s \over d}.
\end{equation}
For a film described by steps all of the same height $s$, the
resistivity is then given by
\begin{equation}
\rho = \rho_{\rm ff} + {1 \over 2\, g_{0}\, \xi}\, {s \over d-s},
\end{equation}
where the factor of $2$ in the denominator accounts for the fact that
the total number of steps $N$ is twice the number of ``up'' and
``down'' step pairs, and $\xi=L/N$ is the average step-correlation
length. In the case where $d \gg s$, the resistivity can be
approximated by
\begin{equation}
\rho \approx \rho_{\rm ff} + {1 \over 2\, g_{0}\, \xi}\, {s \over d}.
\label{rhogall}
\end{equation}

The resistivity difference $\Delta \rho=\rho-\rho_{\rm ff}$ can be
predicted using the equations above and the computational details for
the simulated Ru thin films. For both film orientations simulated,
transport was along $[1 1 \bar{2}  0]$. Consequently, the
relevant number of channels for both films is given by
\begin{equation}
M={8\,w\,d \over \sqrt{3}\, a\, c}.
\end{equation}
The specific ballistic conductance for this transport direction is
\begin{equation}
g_{0} = {2 e^{2} \over h}\, {8 \over \sqrt{3}\, a\, c}.
\end{equation}
Using the conductance quantum and the lattice parameters for Ru
results in $g_{0}=2.998 \times 10^{15}\, \Omega^{-1}m^{-2}$. We use
$\xi=10$ nm as the step correlation length.  However, the step height
for the two films is slightly different. First, we consider $(0001)$
films. In this case, the step height is $s=c/2$. In this case we
obtain for $(0001)$ films,
\begin{equation}
\Delta \rho_{(0001)} = {0.363\, {\rm nm} \over d}\, \mu
\Omega\,{\rm cm}.\label{eq:dra}
\end{equation}
Second, for the $(1\bar{1}00)$ films, the step height is
$s=a/\sqrt{3}$ and hence, in this case,
\begin{equation}
\Delta \rho_{(1\bar{1}00)} = {0.264\, {\rm nm} \over d}\, \mu
\Omega\,{\rm cm}.
\label{eq:drb}
\end{equation}
In short, the theory predicts a resistivity size effect which is very
similar for the two films.

\begin{figure}
\includegraphics[width=\columnwidth]{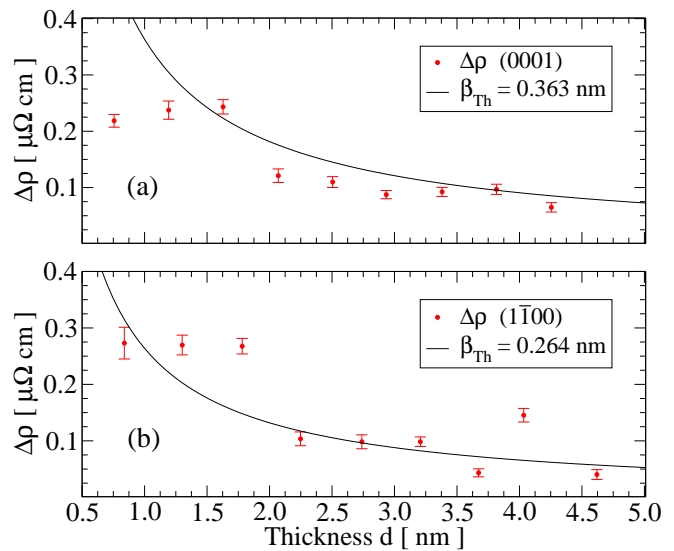}
\caption{$\Delta\rho$ calculated for the (a) $\left(0001\right)$
  surface and for the (b) $\left(1\bar{1}00\right)$
  surface. Additionally, $\Delta\rho_{\text{th}}$ is shown assuming
  the functional form $\Delta\rho = \beta/d$, where $d$ is the film
  thickness and $\beta$ is a prefactor dependent upon the material,
  geometry, and roughness. $\beta_{\text{th}}$ values are the
  theoretical prefactor calculated for the films as found in
  Eqs. \eqref{eq:dra} and \eqref{eq:drb}. The error bars shown 
  reflect the resulting combined error.}
\label{fig:Drho}
\end{figure}

In Fig. \ref{fig:Drho}, the results of this analysis are shown, with
$\Delta\rho$ plotted for both the $\left(0001\right)$ and
$\left(1\bar{1}00\right)$ films. Additionally, the theoretically
expected $\Delta\rho$ as determined in Eqs. \eqref{eq:dra} and
\eqref{eq:drb} via the Landauer formalism have been overlaid on the
data. It can be seen from this plot that the KPM calculated
$\Delta\rho$ follows the same trend and is consistent with the
theoretically predicted $\Delta\rho$ for the examined geometries.


\section{Conclusions}

This paper presents a practical and realistic approach for computation
of the transport properties of perfect-crystal and defective metallic
thin films. Using a realistic TB model parameterized by fitting to DFT
electron energy bands, coupled with the efficient KPM numerical
approach, systems with $10^{6}$ or more sites are easily within
reach. The model for Ru developed here was used to describe the
resistivity size effect due to monolayer-high steps on Ru thin films
with $(0001)$ and $(1 \bar{1}00)$ orientations.

It was found that resistivity increases with decreasing thickness $d$
even in the absence of steps. This is especially evident for the
high-energy $(1\bar{1}00)$ surfaces. The $(1\bar{1}00)$ includes
surface sites with $8$ and $10$ nearest neighbors. By contrast, smooth,
low-energy $(0001)$ surfaces, which are comprised of surface sites
coordinated with $9$ nearest neighbors, display a weaker resistivity
size effect. This potentially suggests a connection between size
effects and surface coordination. Specifically, films with higher
surface energy and lower-coordinated surface sites may tend to exhibit
a measurable size effect even in the absence of surface roughness.
Calculations of the BTE show that these differences in orientation and
surface coordination determine the transport properties of films
without steps. Thus, the results show, for the first time, a
resistivity size effect that does not depend on the presence of
surface defects.

Steps are found to further increase resistivity by an amount $\Delta
\rho$ which also tends to increase with decreasing thickness as $
\Delta \rho \propto 1/d$. This is generally consistent with
observations of the resistivity size effect. However, in comparison
with experimental results, the effect is quite small. This indicates
that other scattering mechanisms are required to explain the observed
resistivity size effect. Some possibilities include surface vacancies,
adatoms, impurities, size-dependent electron-phonon scattering, and
scattering from the substrate. Previous studies of vacancies/adatoms
have shown strong scattering and resistivity increases for a surface
with half of the atoms removed\cite{Zhou:2018aa}. However,
thermodynamically it would be expected that vacancy concentrations are
quite low at room temperature, and it is more likely that adatoms and
vacancies diffuse to form step structures. It is therefore still
uncertain what scattering mechanisms are most likely responsible for
the substantial resistivity size effect seen in metallic thin films.

The small resistivity increases computed for the Ru films appear
generally consistent with the theoretical analysis based on previous
work \cite{Zhou_2018}. Essentially, the very small increase $\Delta
\rho$ with decreasing volume can be traced to the relatively large
specific ballistic conductance $g_{0}=2.998 \times 10^{15}\,
\Omega^{-1} {\rm m}^{-2}$ predicted for Ru. By contrast, the values of
$g_{0}= 9.5 \times 10^{14}\, \Omega^{-1} {\rm m}^{-2}$ for $[100]$
transport and $g_{0}= 8.7 \times 10^{14}\, \Omega^{-1} {\rm m}^{-2}$
for $[110]$ transport were reported for first-principles calculations
of W films \cite{Zheng_2017}. Smaller values of $g_{0}$ have also been
reported for Cu. Using NEGF and DFT, $g_{0} = 0.996 \times 10^{15}\,
\Omega^{-1} {\rm m}^{-2}$ was obtained for Cu
\cite{Zhou_2018}. Similarly, DFT calculations using the Fermi velocity
and the density of states of bulk Cu resulted in $g_{0}=1.10 \times
10^{15}\, \Omega^{-1} {\rm m}^{-2}$ \cite{Schep:1998uy}. Consequently,
Cu and W films would be expected to exhibit a more substantial
resistivity size effect. Indeed, experimental results do apparently
demonstrate much larger increases in W films
\cite{Zhou_2018}. However, it should be noted that the increases seen
in experiments require a much smaller value $g_{0} =1.1-1.4 \times
10^{14}\, \Omega^{-1} {\rm m}^{-2}$ than the value obtained from DFT
calculations. Moreover, experimental results indicate a temperature
dependence which is not explained by Eq. (\ref{rhogall}), although
this effect was possibly due to small errors in the low-temperature
value of $\rho_{\rm ff}$ \cite{Zhou_2018}.
 
In summary, computational results for Ru films show that surface steps
produce a very small resistivity size effect, and experimental results
must be explained by some other mechanism. The behavior found from the
results here are generally consistent with previous theoretical
studies \cite{Zhou_2018}. One important result is the validation that
larger values of the ballistic specific conductivity $g_{0}$ tend to
suppress the resistivity size effect. This observation points to
$g_{0}$ as a critical materials parameter for thin films designed to
mitigate the resistivity size effect. In this respect, Ru appears to
be a very promising material, although further work is needed to
understand the significantly larger size effect seen in experiment.

\acknowledgments

Funding support by the National Science Foundation grants
ECCS-1740228, and the E2CDA-NRI Program of the Semiconductor Research
Corporation under Task 2764.003, as well as support from the Air Force
Office of Scientific Research AFOSR FA9550-19-1-0156 and
FA9550-18-1-0063 is gratefully acknowledged. The authors thank
K. Coffey for illuminating discussions.

\section*{Conflicts of Interest}
The authors have no conflicts to disclose.

\section*{Data Availability Statement}
The data that support the findings of this study are available from the corresponding author upon reasonable request.

\newpage

\appendix
\numberwithin{equation}{section}
\section{Tight-Binding Model Including Fit Parameters}

The tight-binding model described here is based on the general
approach outlined elsewhere \cite{Mehl:1996aa,HbBSES-DAP}.  For
completeness we summarize some of the relevant equations along with
the fit TB parameters for Ru. The starting point for the
parameterization was taken from the original model reported for Ru
\cite{Mehl:1996aa,HbBSES-DAP}. However, in the model developed here,
we assume an orthogonal basis in contrast to the non-orthogonal basis
used in the original model. Consequently, refitting to DFT results was
required. The TB model included 5s, 5p, and 4d orbital states, for a
total of 9 orbitals per Ru site.

The model defines a density function for each atomic site which
depends upon the neighboring sites $j$ within a cutoff radius
$R_c$. For the $i^{\text{th}}$ atomic site, the density $\rho_i$ is
\begin{equation}
\rho_i = \sum_j \exp{\left(-\lambda^2 R_{ij}\right)}
F_{c}\left(R_{ij}\right),
\end{equation}
in which $F_{c}\left(R\right)$ is a smooth cutoff function given by,
\begin{equation}
F_{c}\left(R\right) = \left(1+\exp \left[{R-R_{0} \over l} \right]
\right)^{-1}
\end{equation}
with the values $l=0.5a_{0}$ and $R_{0}=14a_{0}$ ($a_{0}$ is equal to
the Bohr radius). The values for $R_{0}$ and $l$ were kept fixed and
not treated as parameters during the fitting. The on-site energies for
the $\kappa = $ s, p, d orbitals are functions of the density at that
site,
\begin{equation}
h_{\kappa,i} = a_{\kappa} + b_{\kappa}\, \rho_i^{2/3} +
c_{\kappa}\, \rho_i^{4/3}+ d_{\kappa}\, \rho_i^2,
\label{onsiteeq}
\end{equation}
where the $a_{\kappa}, b_{\kappa}, c_{\kappa}$, and $d_{\kappa}$ are
fitting parameters. The two center Slater-Koster hopping integrals are
assumed to take the form of the product between a polynomial and an
exponential function,
\begin{equation}
P_{\gamma}(R) =  \left( e_{\gamma} + f_{\gamma}\,
R\right)\exp{\left( -g_{\gamma}^2 R \right)}
F_C\left(R\right).
\label{hopequation}
\end{equation}
where the $e_{\gamma}$, $f_{\gamma}$, and $g_{\gamma}$ are fitting
parameters and $\gamma$ indicates the type of orbital interaction. The
ten interactions considered for $\gamma$ are: $ss\sigma$, $pp\sigma$,
$sp\sigma$, $dd\sigma$, $sd\sigma$, $pd\sigma$, $pp\pi$, $dd\pi$,
$pd\pi$, and $dd\delta$. For our TB calculations, a cutoff radius of
$R_c=8.731$~\AA\ was used, which corresponds to $16.5a_{0}$. With this
cutoff radius, the on-site density and hopping integrals for each
atomic site included contributions from over 300 nearest neighbors.

In Tables \ref{onsite} and \ref{hops} we list the model parameters
along with the original parameterization for Ru obtained in
Ref. \cite{Mehl:1996aa}.  The parameters corresponding to the choice
of a non-orthogonal basis in Ref. \cite{Mehl:1996aa} are not shown in
Tables \ref{onsite} and \ref{hops}.

  
  \begin{table*}[h]

   \begin{tabular} {| c | c  | c | c | c | c |}
   \hline
   Orbital type & $\lambda$ (Bohr$^{-{1 \over 2}})$ &  a (Ryd) & b (Ryd)& c  (Ryd)& d (Ryd) \\
   \hline
   s & 1.30623255& 0.08931602 & 39.97796548 & 483.111844 & 0.00000000 \\
      &  (1.34252627) &(0.08931602) & (41.20311655) & (1299.430697) & (0.00000000) \\
      \hline
   p & 1.30623255 & 0.67065506 &  33.67551683 & 37.45800567  &  0.00000000\\
      & (1.34252627) & (0.67065506) & (40.89165125) & (36.50451212) & (0.00000000)\\
   \hline
   d & 1.30623255 & 0.06477769 &  2.73143963 & 43.27809398 &  0.00000000\\
      & (1.34252627) & (0.06477769) & (1.15434058) & (40.29400042) & (0.00000000)\\
   \hline
   \end{tabular}
   \caption{Terms for the onsite energies [see Eq. (\ref{onsiteeq})]
     corresponding to different orbital types s, p, and d. For
     comparison, the initial values taken from the non-orthogonal TB
     model are shown in parentheses \cite{Mehl:1996aa}.}
   \label{onsite}
   \end{table*}
  
\begin{table*}
 \begin{tabular}{| c | c | c | c |}
   \hline
 Interaction & $e$ (Ryd) & $f$ (Ryd Bohr$^{-1}$) & $g$ (Bohr$^{-{1 \over 2}}$) \\
    \hline
 ss$\sigma$ & -5.76905601 & -0.53063103 & 0.95191442\\
  & (-10.68255926) & (-0.51916245) & (0.99865171) \\
     \hline
 sp$\sigma$ & 5.54490483 & 0.02408604 & 0.86425581 \\
   & (5.36202489) & (0.02426633) & (0.87998858) \\
     \hline
 sd$\sigma$ & -1.58112292 & -0.09424263 & 0.97033599 \\
   & (-1.63450595) & (-0.09548255) & (0.89393987) \\
     \hline
 pp$\sigma$ & 3.73015359 & 0.02004937 & 0.74717463 \\
   & (3.00392897) & (0.01981091) & (0.72625605) \\
     \hline
 pp$\pi$       & -0.03805903 &  -0.00170700 & 0.54817901\\
   & (-0.08525810) & (-0.00222726) & (0.54121581) \\
     \hline
 pd$\sigma$ & -0.54116349 & 0.00003878 & 0.62107823 \\
   & (-0.50807690) & (0.00061760) & (0.66666094) \\
     \hline
 pd$\pi$        & 0.37232988 & -0.00347231 & 0.68928350\\
   & (0.33314613) & (-0.00386754) & (0.71099612) \\
     \hline
 dd$\sigma$  & -3.15232702 & -0.04735649 & 0.86479684 \\
   & (-2.47448107) & (-0.04632234) & (0.85497356) \\
     \hline
 dd$\pi$ & 3.86881365 & 0.01948502 & 0.93447120 \\
   & (4.17666085) & (0.01943914) & (0.93469295) \\
     \hline
 dd$\delta$ & 5.17396639 & -2.22203525 & 1.12104574 \\
   & (5.28412218) & (-2.14643637) & (1.1103290) \\
     \hline
 \end{tabular} 
 \caption{Parameters for two-center hopping integrals in
   Eq. (\ref{hopequation}) for each interaction type. Starting
   parameters from the original non-orthogonal TB model
   \cite{Mehl:1996aa} are included in parentheses.}
 \label{hops}
 \end{table*}
 
 \newpage

\newpage

\bibliography{metalfilm_2021-PKS-final}
\bibliographystyle{unsrt}

\end{document}